\documentclass[prd,amsmath,amssymb,superscriptaddress,preprintnumbers,twocolumn,10pt]{revtex4-1}

\pdfoutput=1 

\usepackage{graphicx}
\usepackage{dcolumn}
\usepackage{bm}

\usepackage{amssymb}
\usepackage{latexsym}
\usepackage{booktabs}
\usepackage{amsmath}
\usepackage{multirow}
\usepackage{url}

\usepackage{float}
\usepackage[colorlinks=true, linkcolor=red, citecolor=blue]{hyperref}

\usepackage[normalem]{ulem}
\usepackage{color}
\usepackage{array}
\usepackage{enumerate}

\usepackage{lineno}
\usepackage{graphicx}
\usepackage{dcolumn}
\usepackage{bm}
\usepackage{amssymb}
\usepackage{latexsym}
\usepackage{booktabs}
\usepackage{amsmath}
\usepackage{multirow}
\usepackage{url}
\usepackage{footnote}
\usepackage{float}
\usepackage{acro}
\usepackage{hyperref}
\usepackage{cleveref}
\usepackage[section]{placeins}

\DeclareAcronym{SKA}{
   short = SKA ,
   long  = Square Kilometre Array ,
   short-plural =  ,
}

\DeclareAcronym{PBH}{
  short = PBH ,
  long  = primordial black hole ,
  short-plural = s ,
}

\DeclareAcronym{CMB}{
  short = CMB ,
  long  = cosmic microwave background ,
  short-plural =  ,
}

\DeclareAcronym{IGM}{
  short = IGM ,
  long  = intergalactic medium ,
  short-plural =  ,
}

\DeclareAcronym{GW}{
  short = GW ,
  long  = gravitational wave ,
  short-plural = s ,
}
\DeclareAcronym{1D}{
  short = 1D ,
  long  = one-dimensional ,
  short-plural =  ,
}

\DeclareAcronym{MCGs}{
  short = MCGs ,
  long  = molecular-cooling galaxies ,
  short-plural =  ,
}

\DeclareAcronym{ACGs}{
  short = ACGs ,
  long  = atom-cooling galaxies ,
  short-plural =  ,
}

\DeclareAcronym{EoR}{
  short = EoR ,
  long  = epoch of reionization ,
  short-plural =  ,
}

\DeclareAcronym{EDGES}{
  short = EDGES ,
  long  = Experiment To Detect The Global EoR Signature ,
  short-plural =  ,
}

\DeclareAcronym{Fermi-LAT}{
  short = Fermi-LAT ,
  long  = Fermi Large Area Telescope ,
  short-plural =  ,
}

\DeclareAcronym{INTEGRAL}{
  short = INTEGRAL ,
  long  = International Gamma-Ray Astrophysics Laboratory,
  short-plural =  ,
}

\DeclareAcronym{SARAS 3}{
  short = SARAS 3,
  long  = Shaped Antenna measurement of the background Radio Spectrum 3,
  short-plural =  ,
}

\DeclareAcronym{DM}{
   short = DM ,
   long  = dark matter ,
   short-plural =  ,
}

\DeclareAcronym{HEAO}{
   short = HEAO,
   long  = High Energy Astrophysical Observatory ,
   short-plural =  ,
}

\DeclareAcronym{COMPTEL}{
   short = COMPTEL ,
   long  = Imaging Compton Telescope ,
   short-plural =  ,
}

\DeclareAcronym{EGRET}{
   short = EGRET ,
   long  = Energetic Gamma-ray Experiment Telescope ,
   short-plural =  ,
}

\DeclareAcronym{VERITAS}{
   short = VERITAS ,
   long  = Very Energetic Radiation Imaging Telescope Array System ,
   short-plural =  ,
}

\DeclareAcronym{H.E.S.S}{
   short = H.E.S.S.,
   long  = High Energy Stereoscopic System ,
   short-plural =  ,
}

\DeclareAcronym{MAGIC}{
   short = MAGIC,
   long  = Major Atmospheric Gamma Imaging Cherenkov telescopes ,
   short-plural =  ,
}

\DeclareAcronym{LIGO}{
   short = LIGO,
   long  = Laser Interferometer Gravitational-Wave Observatory ,
   short-plural =  ,
}

\DeclareAcronym{VIRGO}{
   short = VIRGO,
   long  = Virgo Gravitational Wave Interferometer  ,
   short-plural =  ,
}

\DeclareAcronym{HERA}{
   short = HERA ,
   long  = Hydrogen Epoch of Reionization Array ,
   short-plural =  ,
}
\usepackage[english]{babel}

\usepackage{etoolbox}
\usepackage{ragged2e}

\AtBeginEnvironment{thebibliography}{\justifying}

\usepackage{amsmath}
\usepackage{graphicx}

\begin{document}


\title{21 cm forest one-dimensional power spectrum as an indirect probe of dark matter particles and primordial black holes}

%

\author{Meng-Lin Zhao}
\affiliation{Liaoning Key Laboratory of Cosmology and Astrophysics, College of Sciences, Northeastern University, Shenyang 110819, China}

\author{Yue Shao}

\affiliation{Department of Physics, Liaoning Normal University, Dalian 116029, China}

\author{Sai Wang}
\affiliation{School of Physics, Hangzhou Normal University, Hangzhou 311121, China}

\author{Xin Zhang}
\thanks{Corresponding author}
\email{zhangxin@neu.edu.cn}
\affiliation{Liaoning Key Laboratory of Cosmology and Astrophysics, College of Sciences, Northeastern University, Shenyang 110819, China}
\affiliation{National Frontiers Science Center for Industrial Intelligence and Systems Optimization, Northeastern University, Shenyang 110819, China}
\affiliation{MOE Key Laboratory of Data Analytics and Optimization for Smart Industry, Northeastern University, Shenyang 110819, China}



\begin{abstract}
Understanding the nature of dark matter (DM) particles remains a pivotal challenge in modern cosmology. Current cosmological research on these phenomena primarily utilizes early-universe cosmic microwave background (CMB) observations and other late-time probes, which predominantly focus on large scales. We introduce a novel probe, the 21 cm forest signal, which can be used to investigate DM properties on small scales during the epoch of reionization, thereby addressing the gap left by other cosmological probes. Annihilation and decay of DM particles, as well as Hawking radiation from PBHs, can heat the intergalactic medium (IGM). This heating suppresses the amplitude of the 21 cm forest 1D power spectrum. Therefore, the 1D power spectrum provides an effective method for constraining DM properties. However, astrophysical heating processes in the early universe can also affect the 21 cm forest 1D power spectrum. In this work, we assess the potential of using the Square Kilometre Array (SKA) to observe the 21 cm forest 1D power spectrum for constraining DM properties, under the assumption that astrophysical heating can be constrained reliably by other independent probes.
Under low astrophysical heating conditions, the 1D power spectrum could constrain the DM annihilation cross section and decay lifetime to $\langle\sigma v\rangle \sim {10^{-31}}\,{\rm cm^{3}\,s^{-1}}$ and $\tau \sim {10^{30}}\,{\rm s}$ for ${10}\,{\rm GeV}$ DM particles, and probe PBHs with masses $\sim {10^{15}}\,{\rm\,g}$ at abundances $f_{\mathrm{PBH}} \simeq 10^{-13}$. These constraints represent improvements of $5$-$6$ orders of magnitude over current limits. Furthermore, the 21 cm forest 1D power spectrum has the potential to exceed existing bounds on sub-GeV DM and to probe PBHs with masses above $10^{18}\,{\rm g}$, which are otherwise inaccessible by conventional cosmological probes. With accumulating observational data and technological advancements, the 21 cm forest emerges as a highly promising tool for probing DM properties.
\end{abstract}

\maketitle


\section{Introduction}\label{sec:intro}

The nature of \ac{DM} remains a pivotal unsolved problem in modern physics.
\ac{DM} comprises over 80\% of the non-relativistic matter content in the universe \cite{Rubakov:2019lyf,ParticleDataGroup:2024cfk}.
Besides direct detection of \ac{DM} particles through particle experiments, cosmological observations provide multiple distinct approaches for indirect \ac{DM} probing \cite{Bertone:2004pz,Gaskins:2016cha,Thorpe-Morgan:2024zcq}.
For massive \ac{DM} particles, they could produce particles of standard model through channels of either annihilation or decay, or both, leading to injection of exotic energy into the \ac{IGM} \cite{Chen:2003gz,Slatyer:2015jla,Slatyer:2015kla,Liu:2019bbm,Liu:2019zez,Nguyen:2025tkl}.
Similarly, \acp{PBH} can inject exotic energy via Hawking radiation \cite{Hawking:1971ei}.
This energy injection heats the \ac{IGM}, elevating its temperature and imprinting distinctive signatures.
Consequently, probing such signatures is crucial for unveiling the nature of \ac{DM}, such as constraining \ac{DM} parameters including the annihilation cross-section, decay lifetime, and \ac{PBH} abundance.

The 21 cm forest signal, manifested as 21 cm absorption lines against high-redshift radio-loud background sources \cite{Furlanetto:2002ng,Carilli:2002ky,Furlanetto:2006dt,Furlanetto:2006jb,Xu:2010us,Xu:2010br,Pritchard:2011xb,Ciardi:2012ik}, provides a promising probe for \ac{DM} \cite{Shimabukuro:2014ava,Shimabukuro:2019gzu,Shimabukuro:2020tbs,Shao:2023agv,Sun:2024ywb,Shimabukuro:2025equ}.
Its absorption depth directly correlates with the hydrogen spin temperature, rendering it exquisitely sensitive to \ac{IGM} temperature variations \cite{Xu:2009dr,Xu:2010br,Soltinsky:2024mzy,Patil:2025fxd}.
Compared to probes such as \ac{CMB}, cosmic rays, and the 21 cm global and power spectra, which are only sensitive to large-scale \ac{DM}-induced heating \cite{Carr:2020gox,Thorpe-Morgan:2024zcq}, the 21 cm forest enables deep exploration of small-scale heating in the early universe.
These large-scale probes are expected to measure averaged heating, inherently neglecting small-scale heating effects.
In contrast, the 21 cm forest signal is sensitive to heating effects on small scales, encoding richer information about \ac{DM}-induced exotic energy injection. 
Furthermore, numerous radio-loud quasars have been observed with well-established abundance models, providing essential background sources for 21 cm forest observations \cite{Haiman:2004ny,Lyu:2023ojm,Banados:2024xds,Niu:2024eyf}. 
Consequently, the 21 cm forest signal is expected to offer superior sensitivity over large-scale probes for constraining \ac{DM} parameters.

The 21 cm forest \ac{1D} power spectrum serves as a crucial observable for extracting signatures of \ac{DM}-induced anomalous energy injection.
Recent studies demonstrate that by measuring scale-dependent correlations in the signal, the \ac{1D} power spectrum significantly enhances the signal-to-noise ratio \cite{Parsons:2011ew,Thyagarajan:2020nch,Shao:2024owi,Shao:2025ohz}, establishing the 21 cm forest as a robust \ac{DM} probe.
The upcoming \ac{SKA}, with its unprecedented sensitivity, will provide an ideal instrument for these observations \cite{Braun:2019gdo}.

In this work, we investigate the prospective sensitivity of the \ac{SKA} in unraveling the nature of \ac{DM} using the \ac{1D} power spectrum of the 21 cm forest.
Taking into account the aforementioned exotic energy injection induced by \ac{DM}, we first simulate the 21 cm forest signal using a multi-scale simulation approach.
We then explore the parameter degeneracy in the spectrum that makes it impossible to distinguish \ac{DM}-induced heating from astrophysical heating with the upcoming \ac{SKA}, expecting to only place upper bounds on the DM-induced heating effects.
In other words, we will obtain upper (or lower) bounds on the \ac{DM} parameters, which compose the main results of this work.
Finally, we demonstrate that the 21 cm forest signal serves as an important probe for both the annihilation and decay of \ac{DM} particles and the Hawking radiation from \acp{PBH}.

The remainder of this paper is organized as follows. In Section \ref{sec:tem}, we summarize the basic theory of exotic energy injection that leads to heating effects on the \ac{IGM} gas. In Section \ref{sec:21 cm}, we introduce the 21 cm forest \ac{1D} power spectrum and its simulations. In Section \ref{sec:Noise}, we briefly demonstrate the Fisher matrix that is used for estimating model parameters. In Section \ref{sec:result},we address the parameter degeneracy and derive projected constraints on the model parameters from the \ac{SKA}. Conclusions are presented in Section \ref{sec:summary}. 
Throughout this paper, we adopt the cosmological parameters from the Planck 2018 results \cite{Planck:2018vyg}.

\section{Exotic energy injection}
\label{sec:tem}

The thermal history of \ac{IGM} can be influenced by the exotic energy produced by the annihilation and decay of \ac{DM} particles, as well as by Hawking radiation from \acp{PBH}. In this section, we summarize the \ac{DM}-induced exotic energy injection and and its effects on \ac{IGM} heating. We also discuss astrophysical processes contributing to \ac{IGM} heating.

\subsection{Annihilation and decay of DM particles}

Processes of the annihilation and decay of \ac{DM} particles produce various Standard Model particles. In this work, we consider three kinds of primary products, i.e., photons, electron-positron pairs, and bottom-anti-bottom quark pairs. Undergoing hadronization process, the primary particles subsequently produce secondary products. We consider the secondary products like photons, electrons, and positrons, due to their dominant role in the energy transfer to the \ac{IGM} gas, whose thermal state is thereby influenced. To simulate relevant processes, we utilize the \texttt{PPPC4DMID} \cite{Cirelli:2010xx} and \texttt{pythia} \cite{Bierlich:2022pfr} codes.

Considering the $s$-wave annihilation channel of \ac{DM} particles, we have the exotic energy injection rate density as follows
\cite{Slatyer:2015jla,Slatyer:2015kla,Liu:2019bbm}
\begin{equation}
    \left(\frac{{\rm d}E}{{\rm d}V{\rm d}t}\right)_{\rm inj,ann} = \rho^{2}_{\rm DM} \mathcal{B}(z) (1+z)^{6} c^2 \frac{\langle \sigma v \rangle}{m_{\chi}}\,,\label{eq:darkmatterani}
\end{equation}
where $\rho_{\mathrm{DM}}$ is the present-day energy density of \ac{DM}, $\mathcal{B}(z)$ the boost factor due to clumping of \ac{DM} \cite{Takahashi:2021pse}, $z$ the cosmological redshift, $\langle \sigma v \rangle$ the thermally-average annihilation cross-section, and $m_{\chi}$ the mass of \ac{DM} particles. 

Considering the two-body decay of \ac{DM} particles, we have the exotic energy injection rate density as follows 
\begin{equation}
    \left(\frac{{\rm d}E}{{\rm d}V{\rm d}t}\right)_{\rm inj,dec} = \rho_{\rm DM} (1+z)^{3} c^2 \frac{1}{\tau}\,, \label{eq:darkmatterdecay} 
\end{equation}
where $\tau$ is the lifetime of \ac{DM} particles. 
In general, only a fraction of the \ac{DM} particles may be capable of annihilation or decay.
This is accounted for by introducing the factors $f_{\rm ann}^{2}$ or $f_{\rm dec}$ on the right-hand side of the above formulas, corresponding to the annihilating or decaying fraction, respectively.
In the present work, we make the simplifying assumption that all \ac{DM} particles can annihilate to or decay into Standard Model particles, which corresponds to setting $f_{\rm ann} = 1$ or $f_{\rm dec} = 1$.

\subsection{Hawking radiation of PBHs}

Due to the mechanism of Hawking radiation, \acp{PBH} produce various Standard Model particles \cite{Hawking:1971ei}, which are called the primary products. Undergo processes such as annihilation, decay, and hadronization, the primary particles produce the secondary products such as photons, electrons, positrons, neutrinos, and others. In this work, we focus on \acp{PBH} in the mass range of $\sim10^{15}-10^{18}\,\mathrm{g}$. For the emission products, we consider the photons and electron-positron pairs, since they are important for the energy transfer to the \ac{IGM} gas. Their spectra, denoted as ${\rm d}^{2} N / ({\rm d} E {\rm d} t)|_{X}$ with $X$ standing for either $\gamma$ or $e^{\pm}$, are calculated via the \texttt{BlackHawk} \cite{Auffinger:2020ztk} code. 
Therefore, we have the exotic energy injection rate density as follows \cite{Mena:2019nhm,Saha:2021pqf,Cang:2021owu}
\begin{eqnarray}
    && \left(\frac{{\rm d}E}{{\rm d}V{\rm d}t}\right)_{\rm inj,PBH} \nonumber\\ && = \int_{0}^{5\rm GeV} \frac{{\rm d}^{2} N}{{\rm d} E {\rm d} t} \Big|_{\rm \gamma} n_{\rm PBH} E {\rm d} E \nonumber\\
    && + \int_{m_{\rm e}c^{2}}^{5\rm GeV} \frac{{\rm d}^{2} N}{{\rm d} E {\rm d} t}\Big|_{\rm e^{\pm}} n_{\rm PBH} (E - m_{\rm e} c^{2}) {\rm d} E\,, \label{eq:Einjtotal}
\end{eqnarray}
where $m_{e}$ is the electron mass, and $n_{\mathrm{PBH}}$ the comoving number density of \acp{PBH}, i.e.,
\begin{equation}
    n_{\rm PBH} = \frac{f_{\rm PBH} \rho_{\rm DM} }{M_{\rm PBH}}\,, \label{eq:npbh}
\end{equation}
with $f_{\mathrm{PBH}}$ being the abundance of \ac{DM} as \acp{PBH} and $M_{\mathrm{PBH}}$ being the mass of \acp{PBH}. In this work, we have assumed a monochromatic mass function of \acp{PBH} for simplicity. However, it can be straightforwardly extended to consider other types of mass function.

\subsection{IGM heating}
\label{subseq:IGM}

The \ac{DM}-induced exotic energy can lead to the \ac{IGM} heating, since it is deposited into the \ac{IGM} gas. In this work, we utilize the delayed deposition, rather than the simultaneous deposition. The energy deposition is calculated using the well-established \texttt{Darkhistory} code, which is dedicated to modeling such processes \cite{Liu:2019bbm}. Hence, the heating rate, denoted as $\epsilon_{\mathrm{exo,h}}$, is given by 
\begin{eqnarray}
    \epsilon_{\rm exo,\rm h} = F_{\rm heat}(z)  \left(\frac{{\rm d} E}{{\rm d} V {\rm d} t}\right)_{\rm inj,L}\,, \label{eq:epsilonheat}
\end{eqnarray}
where $F_{\mathrm{heat}}(z)$ is the energy deposition efficiency through processes of heating, which is robustly calculated by the \texttt{Darkhistory} code based on well-established physical treatments \cite{Liu:2019bbm}.
The subscript $_{\mathrm{L}}$ stands for either the annihilation, or the decay, or the Hawking radiation. 

Besides the \ac{DM}-induced exotic energy, the astrophysical X rays can also lead to the \ac{IGM} heating. The total X-ray emissivity $\epsilon_{\mathrm{X}}$ is given by
\begin{equation}
    \frac{2}{3} \frac{\epsilon_{\rm X}} {k_{\rm B} n_{\rm b}\, H(z)} = 5 \times 10^{4}~f_{\rm X} \left(\frac{f_{\ast}}{0.1} \frac{{\rm d} f_{\rm coll} / {\rm d} z}{0.01} \frac{1+z}{10}\right)\,{\rm K} \,,
\end{equation}
where $n_{b}$ denotes the number density of baryons, $H(z)$ the Hubble parameter at redshift $z$.
$f_{\mathrm{X}}$ is the X-ray productivity in the early universe, which is defined as a global, normalized efficiency factor, with no intrinsic spatial or redshift dependence.
$f_{\ast}$ is the star formation efficiency, and $f_{\mathrm{coll}}$ is the fraction of matter collapsed into halos.
We will discuss $f_{\mathrm{X}}$ in the following section, while $f_{\ast}$ and $f_{\mathrm{coll}}$ are adopted from previous theoretical estimates \cite{Furlanetto:2006jb,Pritchard:2011xb}.
Considering the deposition, we express the heating rate per baryon as follows
\begin{equation}
\epsilon_{\mathrm{X,h}} = f_{\mathrm{heat}} \epsilon_{\mathrm{X}}\,,
\end{equation}
where $f_{\mathrm{heat}}$ is the deposition efficiency, as demonstrated in Refs.~\cite{Furlanetto:2006jb,Pritchard:2011xb}. 

\begin{figure}[htbp]
    \centering
    \includegraphics[width=1.0\linewidth]{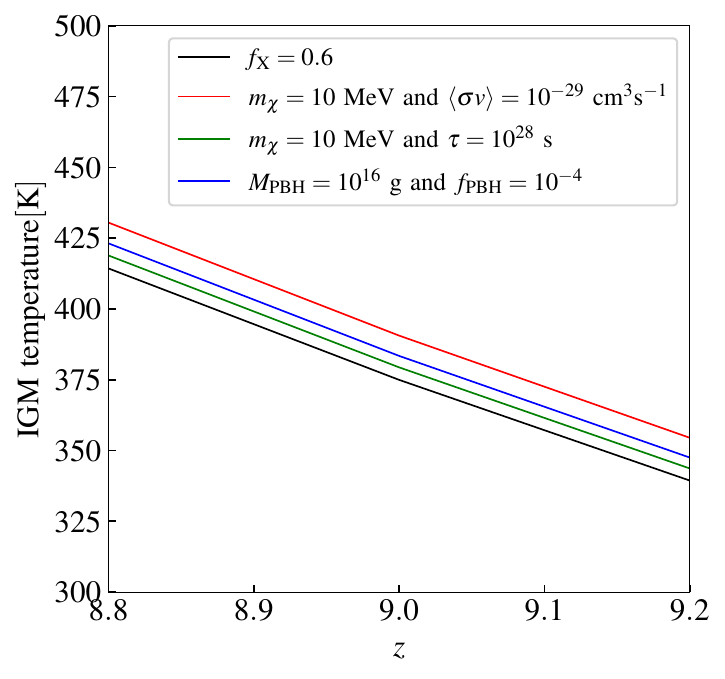}
    \caption{Thermal history of \ac{IGM} under different exotic energy injection scenarios.
    Black curve represents the fiducial value  with $f_{\rm X} = 0.6$.
    Red curve shows the thermal history of \ac{IGM} in the presence of \ac{DM} particle annihilation with $ m_{\chi} = 10 ~\rm{MeV}$  and $\langle \sigma v \rangle = 10^{-29}~ \rm{cm}^{3}s^{-1}$.
    Green curve shows the thermal history of \ac{IGM} in the presence of \ac{DM} particle decay with $ m_{\chi} = 10 ~\rm{MeV}$  and $\tau = 10^{28}~ \rm{s}$. 
    Blue curve shows the thermal history of \ac{IGM} in the presence of PBH with $ M_{\rm PBH} = 10^{16} ~\rm{g}$  and $f_{\rm PBH} = 10^{-4}$.
    }
    \label{fig:Tk}
\end{figure}

Taking into account the heating effects from both astrophysical X rays and \ac{DM}-induced exotic energy, we obtain the evolution equation for the \ac{IGM} temperature, denoted as $T_{\mathrm{K}}$, as follows \cite{Facchinetti:2023slb}
\begin{equation}
\frac{{\rm d} T_{\rm K}}{{\rm d} z} = \frac{2}{3 k_{\rm B} n_{\rm b}} \frac{{\rm d} t}{{\rm d} z} (\epsilon_{\rm exo,\rm h} + \epsilon_{\rm X,\rm h}) + \frac{2 T_{\rm K}}{3 n_{\rm b}} \frac{{\rm d} n_{\rm b}}{{\rm d} z} \,,\label{eq:TK}
\end{equation}
where we have ${\rm d}t/{\rm d}z=1/[H(z)(1+z)]$ with $t$ being the cosmic time corresponding to $z$. In order to visualize these heating effects, we depict $T_{\mathrm{K}}$ as a function of $z$ in Fig.~\ref{fig:Tk}. 

\section{The 21 cm forest signal}
\label{sec:21 cm}

In this section, we first introduce the 1D power spectrum as a key statistical tool for characterizing the 21 cm forest signal. After establishing this theoretical framework, we then present our multi-scale simulation approach to model this signal, following the methodology developed in Refs.~\cite{Shao:2023agv,Sun:2024ywb}.

\subsection{The 21 cm forest 1D power spectrum}

The differential brightness temperature of the 21 cm forest at $z$ is expressed as 
\cite{Furlanetto:2006jb,Pritchard:2011xb}
\begin{eqnarray}
    \delta T_{\rm b}(z) &&\approx 0.0085~ x_{\rm HI}(z) [ 1 + \delta(z)](1+z)^{1/2} \nonumber\\&&\times \left[ \frac{T_{\rm S}(z) - T_{\rm point}(z)}{T_{\rm S}(z)} \right] \nonumber\\&&\times \left(\frac{\Omega_{\rm b} h^{2}}{0.22} \frac{0.14}{\Omega_{\rm m} h^{2}} \right)\, {\rm K}\,, \label{eq:brightness_tem}
\end{eqnarray}
where $x_{\mathrm{HI}}$ is the neutral fraction of hydrogen, $\delta$ the gas overdensity, $T_{\mathrm{S}}$ the spin temperature, and $T_{\mathrm{point}}$ the brightness temperature of background radio sources. 
Since the 21 cm forest originates from neutral regions, we let $x_{\mathrm{HI}}=1$. 
We will conduct multi-scale simulations of $\delta(z)$ and $T_{\mathrm{S}}(z)$ in the following subsection. 
In \cref{eq:brightness_tem}, $T_{\mathrm{S}}$ is given by
\begin{equation}
    T_{\rm S}^{-1} = \frac{T_{\rm point}^{-1} + x_{\alpha}T_{\alpha}^{-1} + x_{\rm c}T_{\rm K}^{-1}}{1 + x_{\alpha} + x_{\rm c}}\,, \label{eq:TS}
\end{equation} 
where $T_{\alpha}$ is the color temperature of the Lyman-$\alpha$ photons with $x_{\alpha}$ being the Lyman-${\alpha}$ coupling coefficient due to the Wouthuysen-Field effect \cite{Furlanetto:2006jb}, and $T_{\mathrm{K}}$ the kinetic temperature of the \ac{IGM} gas with $x_{\rm c}$ being the coupling coefficient due to collisions between two hydrogen atoms, hydrogen atoms and electrons, as well as hydrogen atoms and protons \cite{Furlanetto:2006jb}.
The corresponding coupling coefficients can be calculated theoretically following Refs.~\cite{Furlanetto:2006jb,Pritchard:2011xb}.
In the mainstream \texttt{21cmFAST} code these coefficients are also estimated following those processes \cite{Mesinger:2010ne}.
Due to the frequent scattering, the color temperature is tightly coupled to the kinetic temperature, i.e., $T_{\alpha}\simeq T_{\rm K}$.
While $T_{\rm point}$ is generally far above the $T_{\rm K}$, therefore, $T_{\rm S}$ is dominated by $T_{\rm K}$.
Also in \cref{eq:brightness_tem}, $T_{\mathrm{point}}$ is given by
\begin{equation}
    T_{\rm point}(z) = (1+z) \frac{c^{2}}{2k_{\rm B}\nu^{2}} \frac{S_{\rm point}(\nu)}{\Omega} \,,\label{eq:radation_tem}
\end{equation}
where $\nu$ is the redshifted frequency of 21 cm photons, $S_{\rm point}$ the flux density of background radio sources, and $\Omega = \pi(\theta/2)^2$ the solid angle of the telescope's beam. Here, $\theta = 1.22 \lambda/D$ is the telescope's angular resolution with $\lambda$ being the redshifted wavelength of 21 cm photons and $D$ the telescope's maximum baseline. 
We take $S_{\rm point}$ as a power-law spectrum parameterized at $\nu_{150} = 150 {\rm ~MHz}$, i.e., \cite{Thyagarajan:2020nch}
\begin{equation}
    S_{\rm point}(\nu) = S_{150}\left(\frac{\nu}{\nu_{150}}\right)^\zeta\,,
\end{equation}
where $\zeta = -1.05$ denotes the spectral index \cite{Carilli:2002ky}, and $S_{150}$ is the source flux density at $150~ {\rm MHz}$, which in this work is $10\,\rm{mJy}$.

The \ac{1D} power spectrum of the 21 cm forest along the line of sight is defined as \cite{Parsons:2011ew} 
\begin{equation}
    P(k) = \lvert \tilde{\delta T}(k)\rvert^{2}\left(\frac{1}{\Delta r} \right) \,, \label{eq:1d_power_spectrum}
\end{equation}
where $\Delta r$ is the comoving length of the line of sight used for calculating the auto-correlation, and $k$ the comoving wavenumber.
Here, $\delta \tilde{T}(k)$ is the Fourier transform of $\delta T_{\rm b}(r)$, i.e.,
\begin{equation}
    \delta \tilde{T}(k) = \int_{r_{s}-\Delta r}^{r_{s}} \delta T_{\rm b}(r)e^{-ikr} {\rm d} r \,,
\end{equation}
where $r_{s}=r(z_{s})$ denotes the comoving distance of background radio sources at $z=z_{s}$, and $\delta T_{\rm b}(r)$ is determined by $\delta T_{\rm b}(z)$ since the comoving distance $r$ is determined by $z$.

\subsection{Simulations}

We adopt the multi-scale simulation approach presented in Refs. \cite{Shao:2023agv,Sun:2024ywb} to model the 21 cm forest signal.
This method connects large-scale \texttt{21cmFAST} \cite{Mesinger:2010ne} simulations with small-scale modeling of low-mass \ac{DM} halos via the conditional halo mass function.
Specifically, this method first generates large-scale density fields, then feeds these density fields into the conditional halo mass function as the initial condition for small-scale simulation.
For computational simplicity, in this work we instead use the cosmic mean density as the initial condition for the conditional halo mass function.
While this simplification discards the large-scale fluctuations captured by \texttt{21cmFAST}, it still retains the small-scale information that the 21 cm forest signal primarily depends.



For the small-scale simulation, we first use the conditional halo mass function to compute the abundance and spatial distribution of low-mass halos, with the conditional halo mass function defined as \cite{Press:1973iz,Cooray:2002dia,Zentner:2006vw}
\begin{equation}
    \begin{split}
\frac{{\rm d} n(M|\delta_0,M_0;z)}{{\rm d} M}
&=\sqrt{\frac{1}{2 \pi}} \frac{\bar{\rho}_{\rm m0} (1+\delta_0)}{M} \left|\frac{{\rm d}\sigma^2(M)}{{\rm d} M}  \right| \\& \times
\frac{\delta_{\rm c}(z) - \delta_0}{[\sigma^2(M)-\sigma^2(M_0)]^{3/2}}\ \\& \times \exp{\left\{-\frac{[\delta_{\rm c}(z) - \delta_0]^2}{2[\sigma^2(M)-\sigma^2(M_0)]}\right\}} \,, 
    \label{eq.HMF_CDM}
    \end{split}
\end{equation}
where $\delta_0$ and $M_0$ represent the overdensity and mass in the simulated grid, respectively. 
$\bar{\rho}_{\rm m0}$ denotes the average density of matter today.
$\sigma^2(M)$ is the variance on mass scale $M$, and $\sigma^2(M_0)$ is the variance of $M_{0}$.
$\delta_{\rm c}(z)$ is the critical overdensity for collapse at redshift $z$, extrapolated to the present time.

We set the lower limit of the halo mass to $10^5 M_{\odot}$, consistent with the revised simulation parameters in Ref.~\cite{Sun:2024ywb}, and take the upper limit to be the halo mass corresponding to a virial temperature $T_{\rm vir}=10^4 {\rm K}$, assuming that halos below this mass do not host the first star-forming galaxies capable of producing significant ionization.
As shown by eq. (\ref{eq:brightness_tem}), the 21 cm brightness temperature is governed mainly by the gas density, the hydrogen neutral fraction, and the spin temperature.
We therefore model both inside and outside density, temperature, and ionization profiles for each halo.
The density profile is assumed to follow an NFW distribution, with gas tracing the \ac{DM} \cite{Navarro:1995iw,Navarro:1996gj,Makino:1997dv,Barkana:2002bm}.
The gas temperature inside the halo is set to its virial temperature, while the outside temperature is determined by the competition between adiabatic cooling and heating.
The heating process includes X-rays produced by astrophysical processes, as well as energy injection caused by \ac{DM} processes (including \ac{DM} particle annihilation, decay, and Hawking radiation from primordial black holes).

The volume of the grid used in the small-scale simulation is $(2\, {\rm Mpc})^3$.
In each grid, we calculate the number of \ac{DM} halos within different mass intervals according to the conditional mass function and randomly distribute them.
In our simulation, each grid is divided into $500^3$ voxels, corresponding to a volume of $(4\, {\rm kpc})^3$.
The density, temperature, and ionization fraction within each voxel are determined using the methods described above.
To obtain a longer 21 cm forest signal, we concatenate five such grids to form a $2\, {\rm Mpc}\,\times\,2\, {\rm Mpc}\,\times \,10\, {\rm Mpc} $ volume.
This assembled grid comprises 500 × 500 independent sightlines, each spanning a length of 10 Mpc.
To simulate the multiple neutral segments that can be intercepted in a single radio-loud quasar spectrum, we select 10 of these sightlines for analysis \cite{Shao:2023agv}.

We conducted 10 such simulations at redshift 9, with each simulation representing the 21 cm forest spectrum associated with a radio-loud quasar.
In every simulation we selected 10 lines of sight, yielding a total of 100 individual 21 cm forest spectra and their corresponding 1D power spectra. Finally, we averaged these 100 power spectra to obtain the final mean 1D power spectrum.

Furthermore, astrophysical processes, such as X-rays from the first galaxies, can heat the \ac{IGM} and create degeneracies  \ac{DM} processes.
To constrain the \ac{DM} parameters, we must assign a prior to the X-ray productivity $f_{\rm X}$.
Observations from the \ac{HERA} indicate that the temperature of \ac{IGM} at redshift $z \sim 8$ is constrained within the range $15.6 {\rm\, K} < T_{\rm K}<656.7 {\rm \,K}$ \cite{HERA:2022wmy}, and the corresponding X-ray productivity $f_{\rm X}$ ranges from $0.02$ to $0.6$ \cite{Shao:2024owi}. 
Although this estimate only accounts for X-ray heating of the IGM and ignores contributions from other heating processes, we adopt \( f_{\rm X}=0.02 \) and \( f_{\rm X}=0.6 \) as priors to represent scenarios of weak and strong X-ray heating, respectively.

Furthermore, in the present work, we use the Fisher information matrix to constrain the nature of \ac{DM}.
The Fisher matrix requires specific fiducial values for these parameters.
As discussed, we select \( f_{\rm X}=0.02 \) and \( f_{\rm X}=0.6 \) as two representative fiducial values.
The resulting constraints and degeneracies are shown in Figs.~\ref{fig:ann_correlation_05} and \ref{fig:ann_correlation_002}.
For instance, Fig.~\ref{fig:ann_correlation_05} shows the degeneracies between \( f_{\rm X} \) and \ac{DM} parameters, under the assumption of the fiducial value of \( f_{\rm X} = 0.6 \). The blue shaded regions show the forecasted \(1\sigma\) uncertainties for \( f_{\rm X} \) and the \ac{DM} parameters.
We here show typical cases of \ac{DM} annihilation, decay, and \ac{PBH} Hawking radiation. A more complete analysis would involve showing how the \(1\sigma\) errors on other parameters vary with \( f_{\rm X} \).
However, generating 21 cm forest simulations is very time-consuming.
Given this constraint, we are not able to estimate relevant parameter errors for all possible values of $f_{\rm X}$ in this work.
Therefore, we only estimate the parameter errors at these two endpoint values of $f_{\rm X}$ as representative cases.

\section{Fisher matrix}
\label{sec:Noise}
We utilize the Fisher information matrix to estimate the capability of the \ac{SKA} to constrain \ac{DM} parameters via the 21 cm forest \ac{1D} power spectrum.

\begin{figure}
    \centering
    \includegraphics[width=1\linewidth]{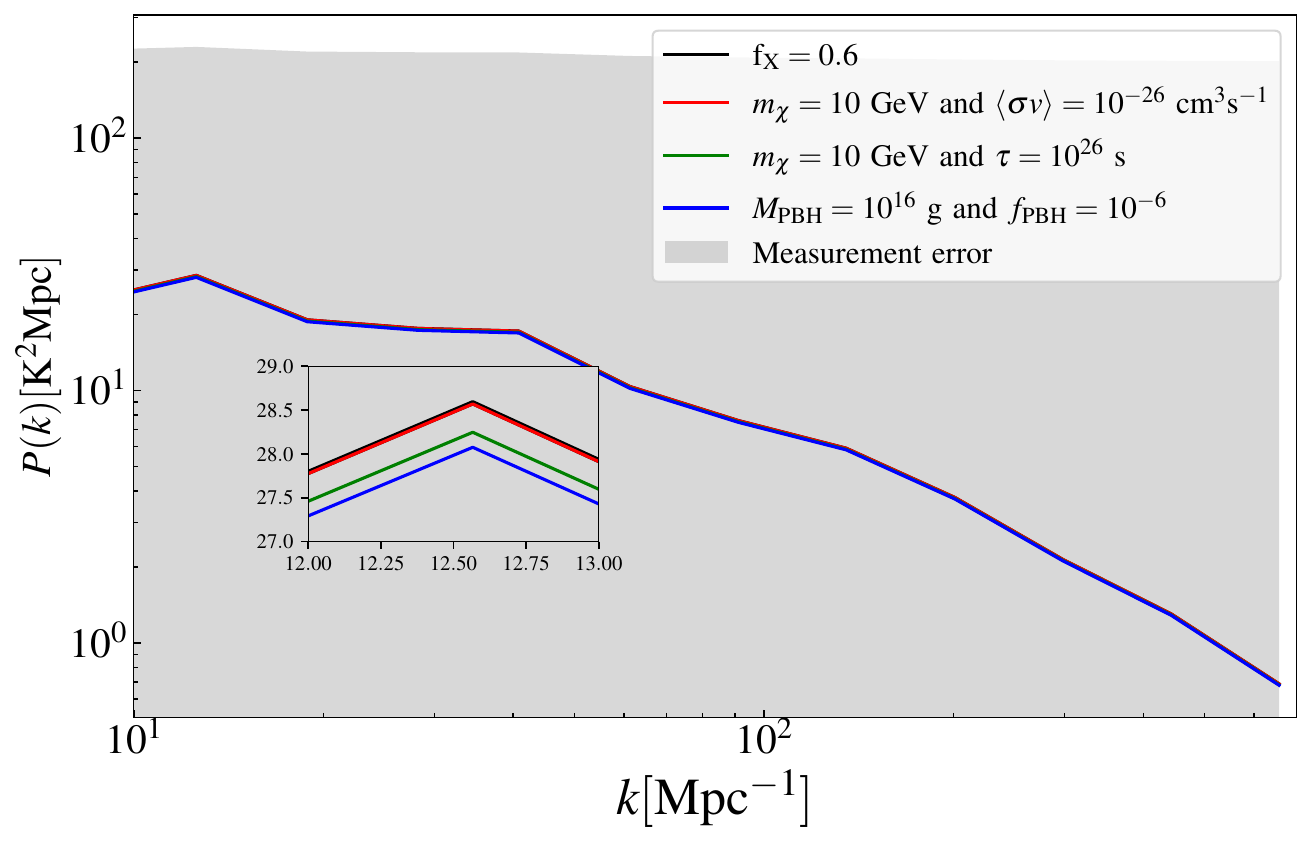}
    \includegraphics[width=1\linewidth]{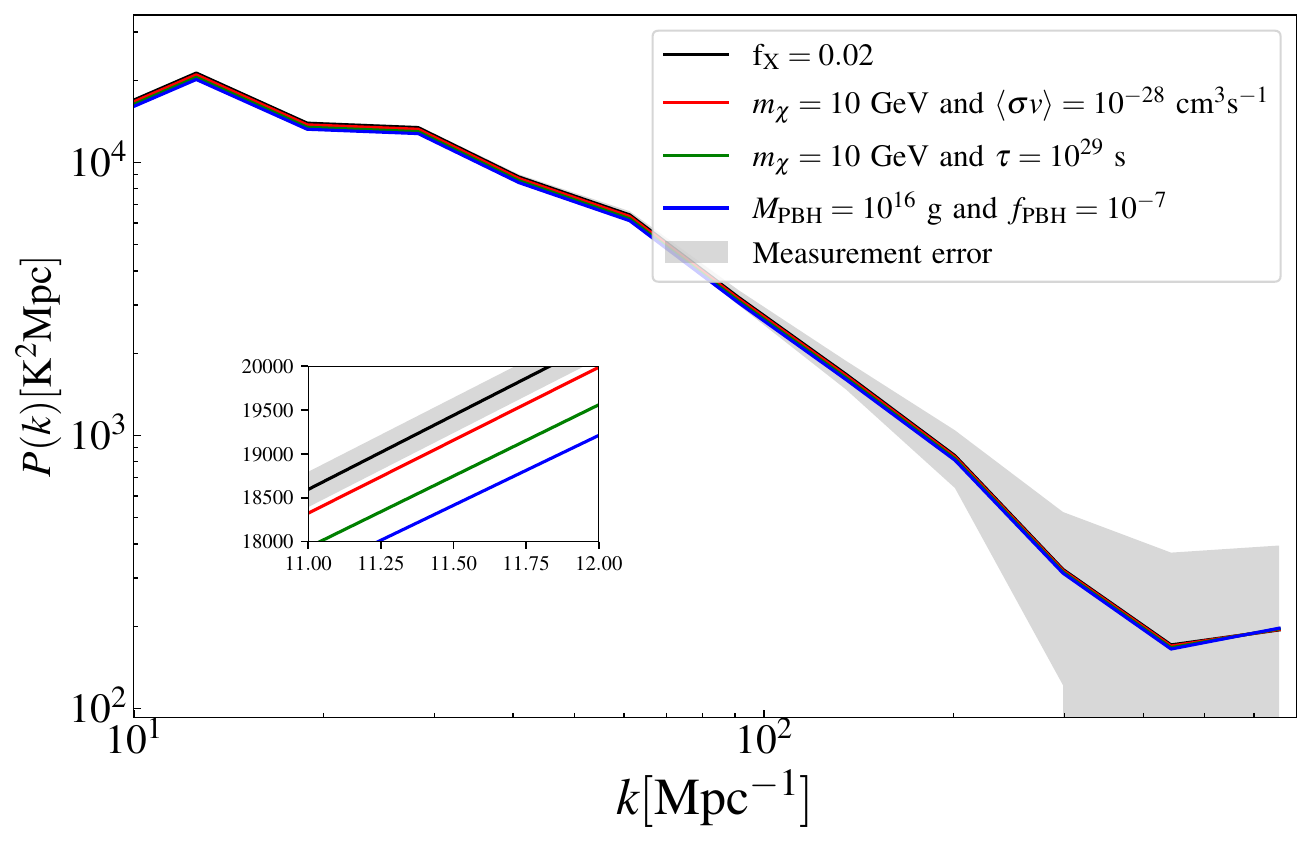}
    
    \caption{1D power spectrum under different scenarios.
    The measurement errors are shown by the gray regions.    
    Black curves show the fiducial model of the 1D power spectrum heated by only astrophysical X-ray sources with $f_{\rm X} = 0.6$ (upper panel) and $f_{\rm X} = 0.02$ (lower panel).
    The red curves correspond to models including \ac{DM} particle annihilation with $ m_{\chi} = 10 ~\rm{GeV}$, and $\langle \sigma v \rangle = 10^{-26}~ \rm{cm}^{3}s^{-1}$ (upper panel) and  $\langle \sigma v \rangle = 10^{-28}~ \rm{cm}^{3}s^{-1}$ (lower panel).
    The green curves correspond to models including \ac{DM} particle decay with $ m_{\chi} = 10 ~\rm{GeV}$ and $\tau = 10^{26}~ \rm{s}$ (upper panel) and $\tau = 10^{29}~ \rm{s}$ (lower panel). 
    The blue curves correspond to models including \ac{PBH} with $ M_{\rm PBH} = 10^{16} ~\rm{g}$ and $f_{\rm PBH} = 10^{-6}$ (upper panel) and $f_{\rm PBH} = 10^{-7}$ (lower panel).}
    \label{fig:error}
\end{figure}

Our Fisher information matrix takes the form 
\begin{equation}
    F_{ij} = \sum_{l}^{N_k} \frac{1}{\sigma_{\rm tot}(k_{l})}\frac{\partial P(k_{l})} {\partial \theta_i} \frac{\partial P(k_{l})} {\partial \theta_j}\ . 
    \label{eq:FisherMatrix_forest}
\end{equation}
Here $N_{k}$ represents the total number of $k$ bins, given by $N_{m} \times N_{s}$.
$N_{m}$ is the number of $k$ bins in each segment, and $N_s$ is the number of neutral segments.
$N_{m}$ is adopted following Ref.~\cite{Shao:2023agv}, and $N_{s}$ is $100$.
$\sigma_{\rm tot}(k_{l})$ represents the total noise for the 21 cm forest \ac{1D} power spectrum in wavenumber bin $k_{l}$.
$\theta_{i}$ and $\theta_{j}$ are the $i$-th and $j$-th parameters in the parameter set.

The total noise mainly araises from two sources, which are given by
\begin{equation}
    \sigma_{\rm tot}(k_{l}) \equiv \sigma_{{\rm ins}} + \sigma_{\rm sam}(k_{l}) \ , \label{eq:noise}
\end{equation}
where $\sigma_{{\rm ins}}$ is the instrumental noise and $\sigma_{\rm sam}(k_{l})$ the sample variance.
The instrumental noise is
\begin{equation}
    \sigma_{{\rm ins}} = \frac{1}{\sqrt{N_{s}}} \left(\frac{\lambda_{z}^{2} T_{{\rm sys}}}{ A_{{\rm eff}} \Omega}\right)^{2}
\left(\frac{\Delta r}{\delta \nu \delta t }\right) \ ,
\end{equation}
where $\lambda_{z}$ denotes the observational wavelength of 21 cm photons at redshift $z$, $T_{\rm sys}$ denotes the system temperature and $A_{\rm eff}$ the effective area.
The solid angle of telescope beam is denoted by $\Omega$.
$\delta \nu$ and $\delta t$ are the bandwidth and integration time for each source, respectively.
In this work, we adopt $A_{\rm eff}/T_{\rm sys} = 538.4 {\rm~m}^2 {\rm~K}^{-1}$ \cite{Braun:2019gdo}, $\delta \nu = 0.56 {\rm~MHz}$, and $\delta t = 100 {\rm~h}$.
The sample variance on the \ac{1D} power spectrum is 
\begin{equation}
    \sigma_{\rm sam}(k)=\frac{\sigma_{\rm 1D}(k)}{\sqrt{N_{k}}} \ ,
\end{equation}
where $\sigma_{\rm 1D}(k)$ represents the standard deviation of 1D power spectrum, which we estimate following Ref.~\cite{Shao:2023agv}.

The marginalized uncertainty $\sigma_{\theta_{i}}$ for a given parameter $\theta_{i}$ satisfies $\sigma_{\theta_{i}} \geq \sqrt{(F^{-1})_{ii}}$ \cite{Aitken_Silverstone_1942}, indicating that the Fisher matrix provides a conservative estimate of the parameter uncertainties.
This conclusion holds under the ideal assumption of negligible systematic effects, such as those from systematic effects and radio frequency interference.
In practice, the presence of such systematics would degrade the experimental sensitivity.
The $1\sigma$ uncertainty of the parameter is its standard deviation.
Moreover, the inverse of the Fisher matrix is the covariance matrix $C_{ij}$, which quantifies the correlations between parameters.
The correlation between parameters $\theta_{i}$ and $\theta_{j}$ is given by the dimensionless correlation coefficient $R_{ij}$, which is defined as $R_{ij} = C_{ij}/\sqrt{ C_{ii} C_{jj} }$.

We demonstrate how the 21 cm forest \ac{1D} spectrum responds to exotic energy injections and present the \ac{SKA}'s measurement errors on the \ac{1D} power spectrum, as shown in Fig. \ref{fig:error}.
Fig. \ref{fig:error} indicates that under conditions of intense astrophysical X-ray heating, the amplitude of the 21 cm forest \ac{1D} power spectrum is significantly suppressed, making it difficult to detect.
Utilizing the 21 cm forest \ac{1D} power spectrum for \ac{DM} detection is therefore challenging in such scenarios.
Conversely, when X-ray heating is relatively weak, the 21 cm forest \ac{1D} power spectrum signal is mainly influenced by \ac{DM}.
Therefore, in this case, we can better explore \ac{DM} properties using this signal.

\section{Results and discussion}
\label{sec:result}

\begin{figure*}[!htbp]
    \centering
\includegraphics[width=0.3\linewidth]{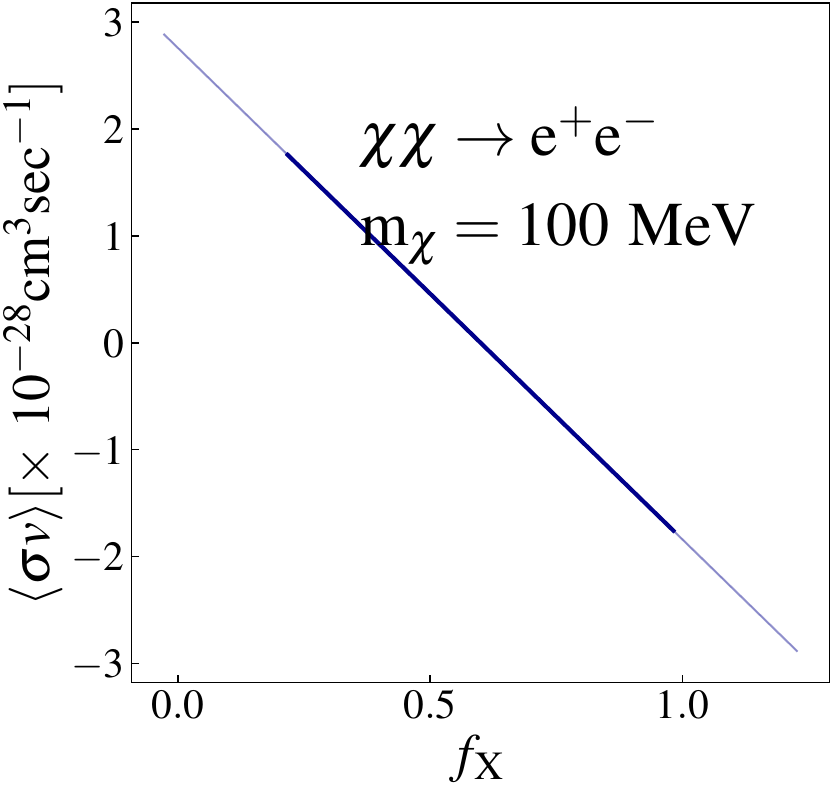} 
    \includegraphics[width=0.33\linewidth]{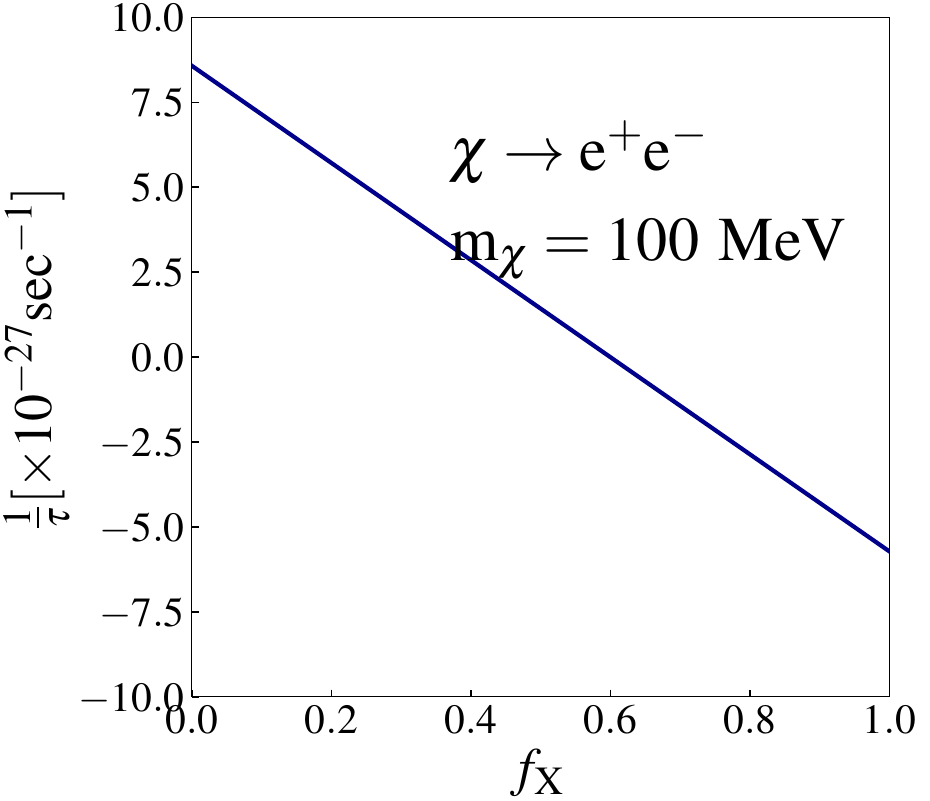}    
    \includegraphics[width=0.33\linewidth]{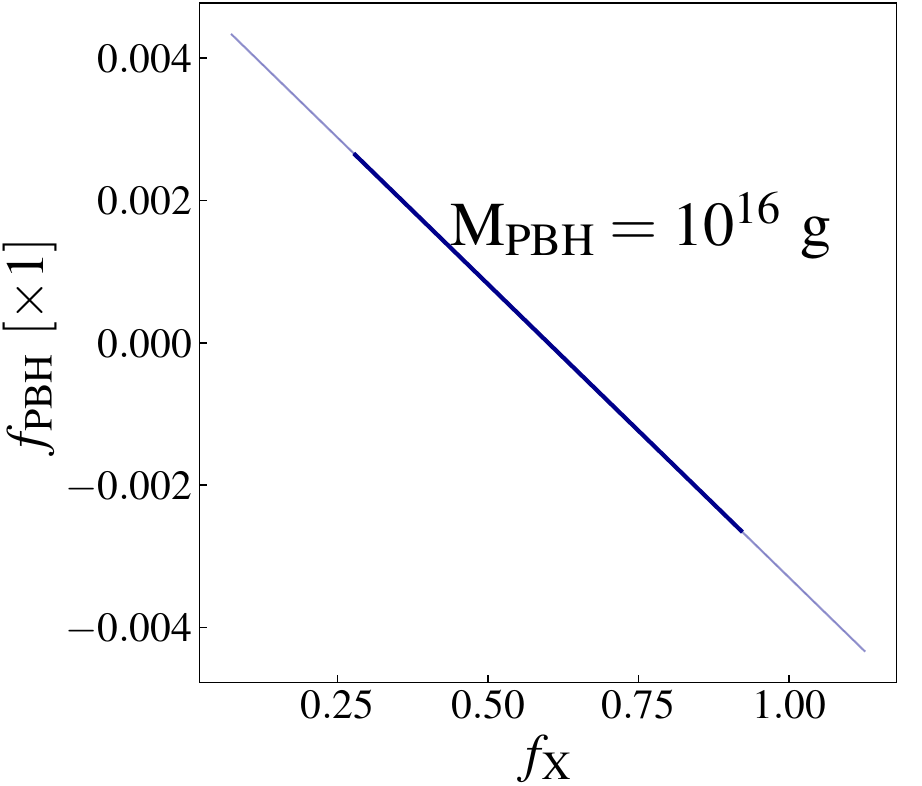}
    \caption{Constraints on $f_{\rm X}$ and \ac{DM} parameters using 21 cm forest 1D power spectrum.
   Left, middle, and right panels show the results for \ac{DM} annihilation, decay, and \ac{PBH} Hawking radiation.
    The fiducial value of $f_{\rm X}$ is 0.6, while \ac{DM} parameter's fiducial values are set to 0.  
    The dark and light regions represent the $1 \sigma$ and $2 \sigma$ confidence contours, respectively.}
    \label{fig:ann_correlation_05}
\end{figure*}

\begin{figure*}[!htbp]
    \centering
\includegraphics[width=0.32\linewidth]{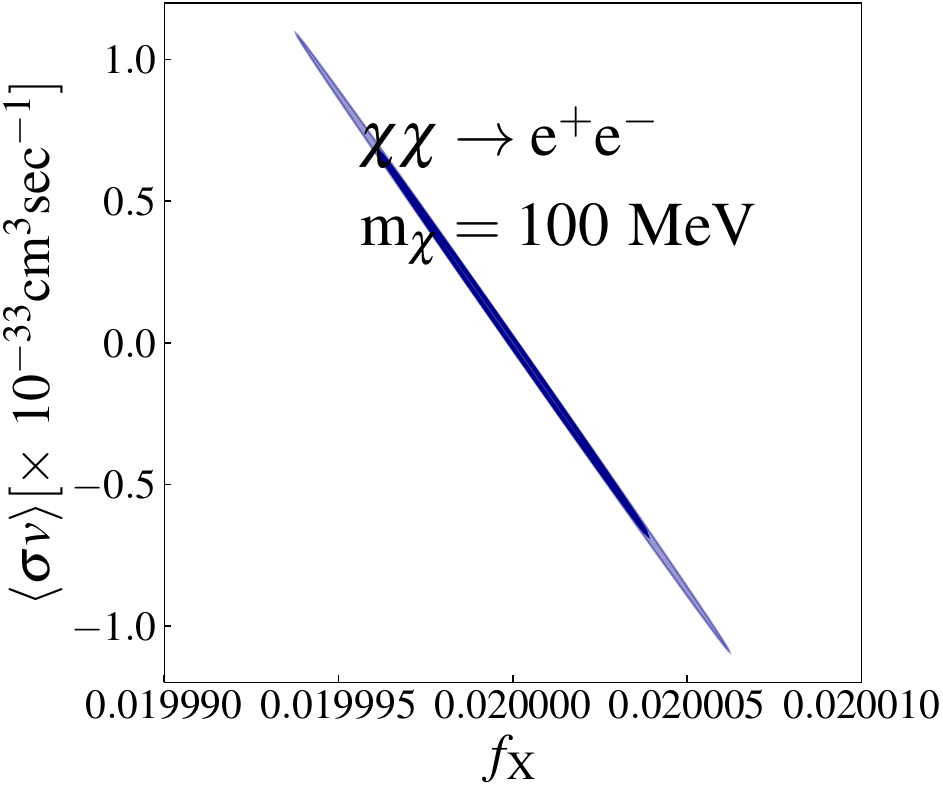} 
    \includegraphics[width=0.32\linewidth]{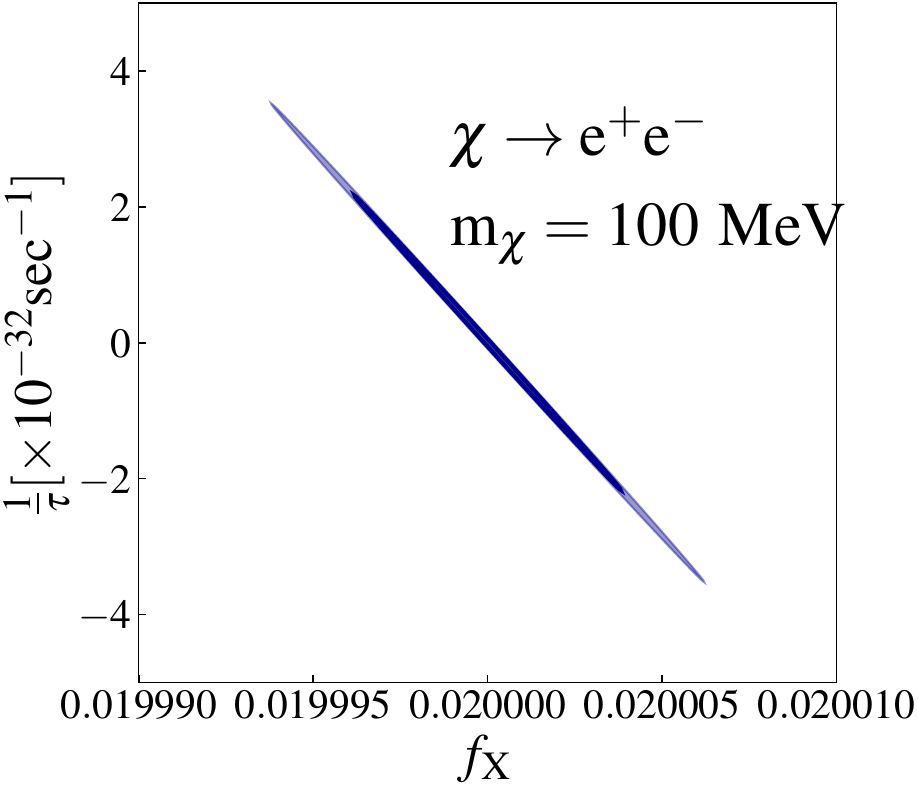}    
    \includegraphics[width=0.32\linewidth]{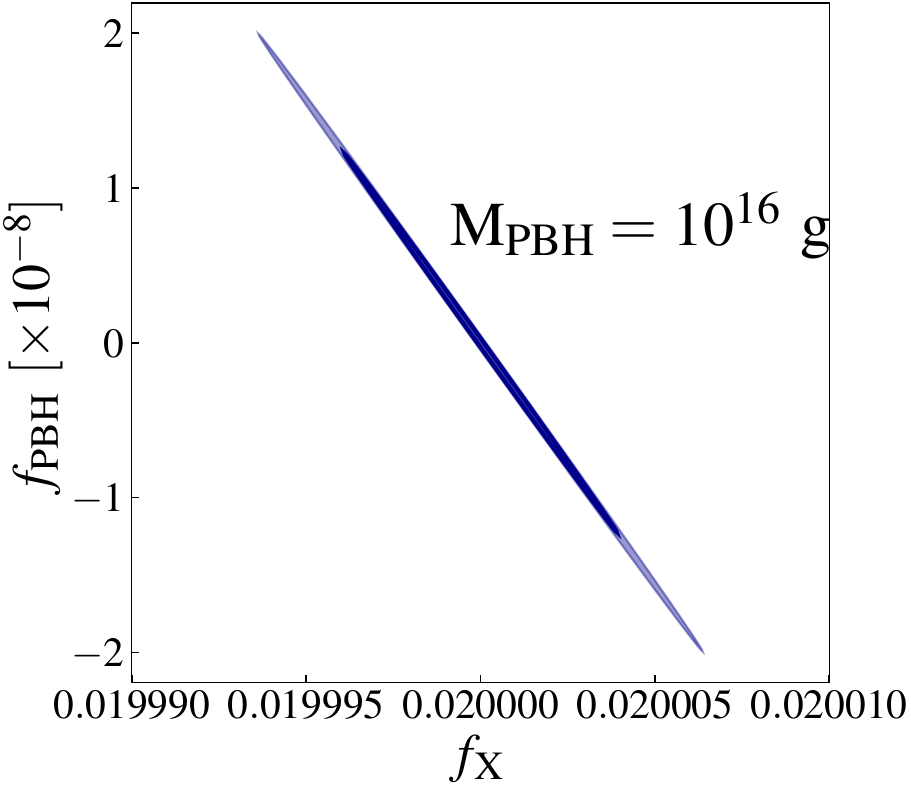}
    \caption{Constraints on $f_{\rm X}$ and \ac{DM} parameters using 21 cm forest 1D power spectrum.
   Left, middle, and right panels show the results for \ac{DM} annihilation, decay, and \ac{PBH} Hawking radiation.
    The fiducial value of $f_{\rm X}$ is 0.02, while \ac{DM} parameter's fiducial values are set to 0.  
    The dark and light regions represent the $1 \sigma$ and $2 \sigma$ confidence contours, respectively.}
    \label{fig:ann_correlation_002}
\end{figure*}

\begin{figure*}[!htbp]
    \centering
    \includegraphics[width=0.32\linewidth]{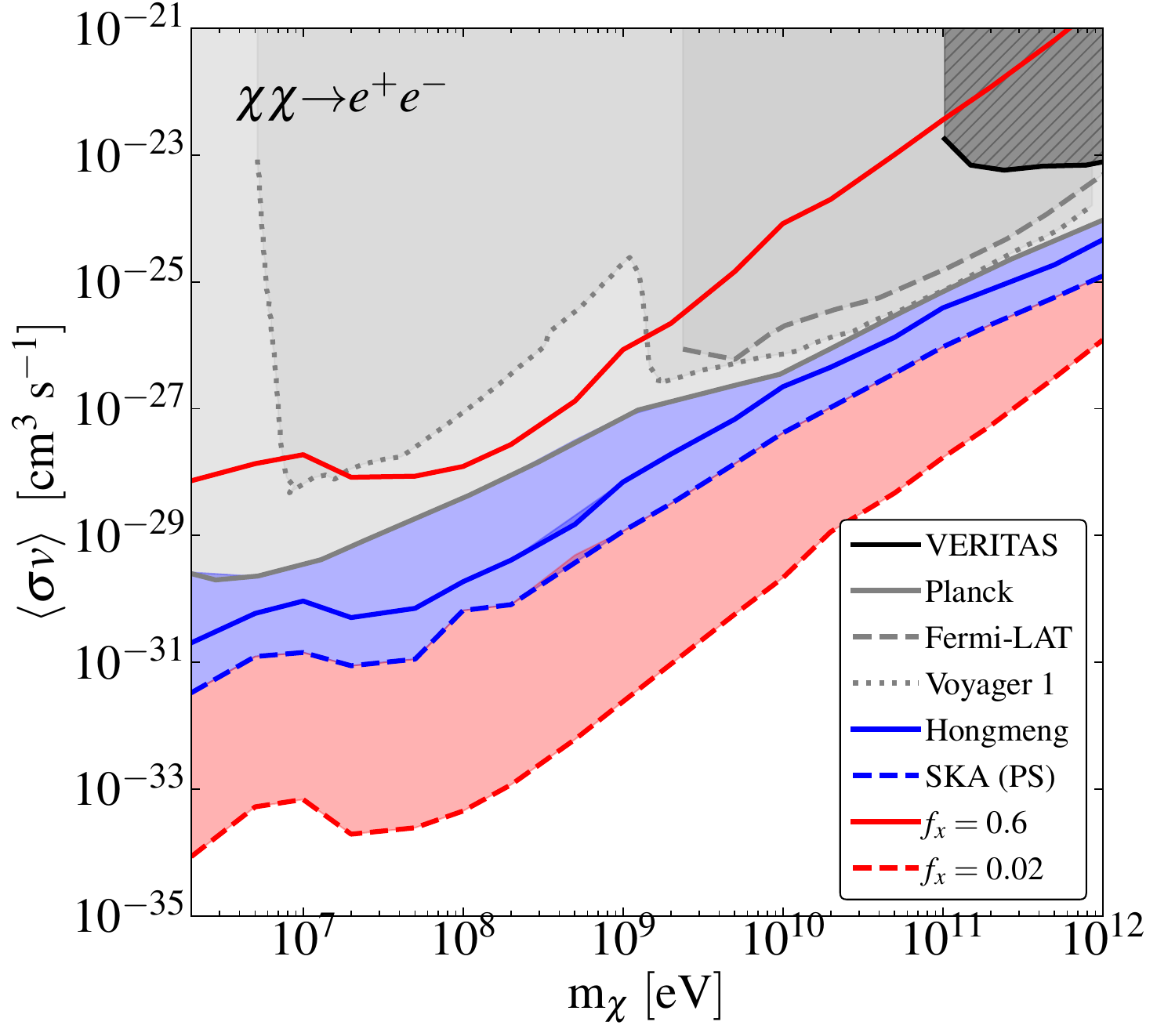}
    \includegraphics[width=0.32\linewidth]{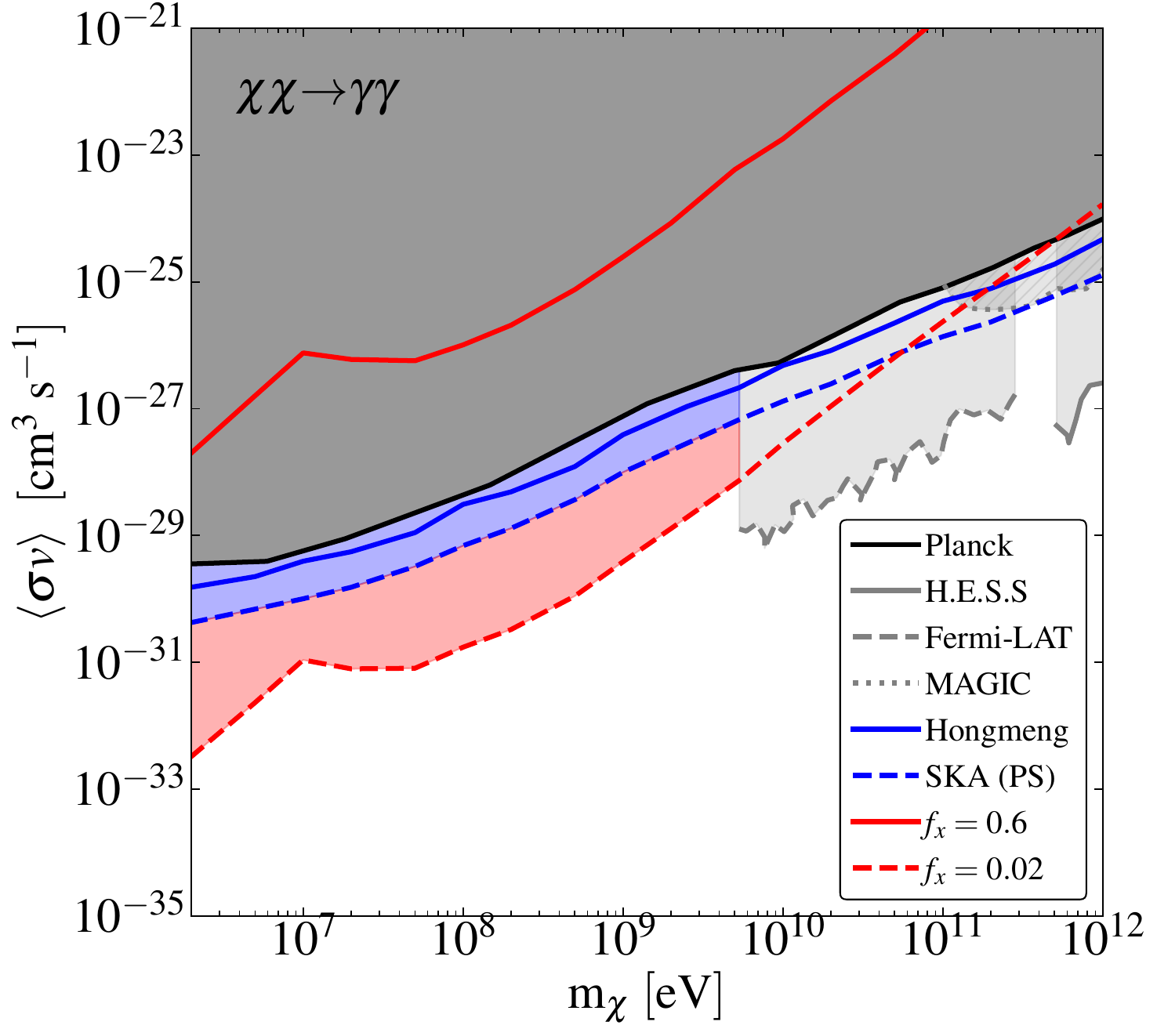}
    \includegraphics[width=0.32\linewidth]{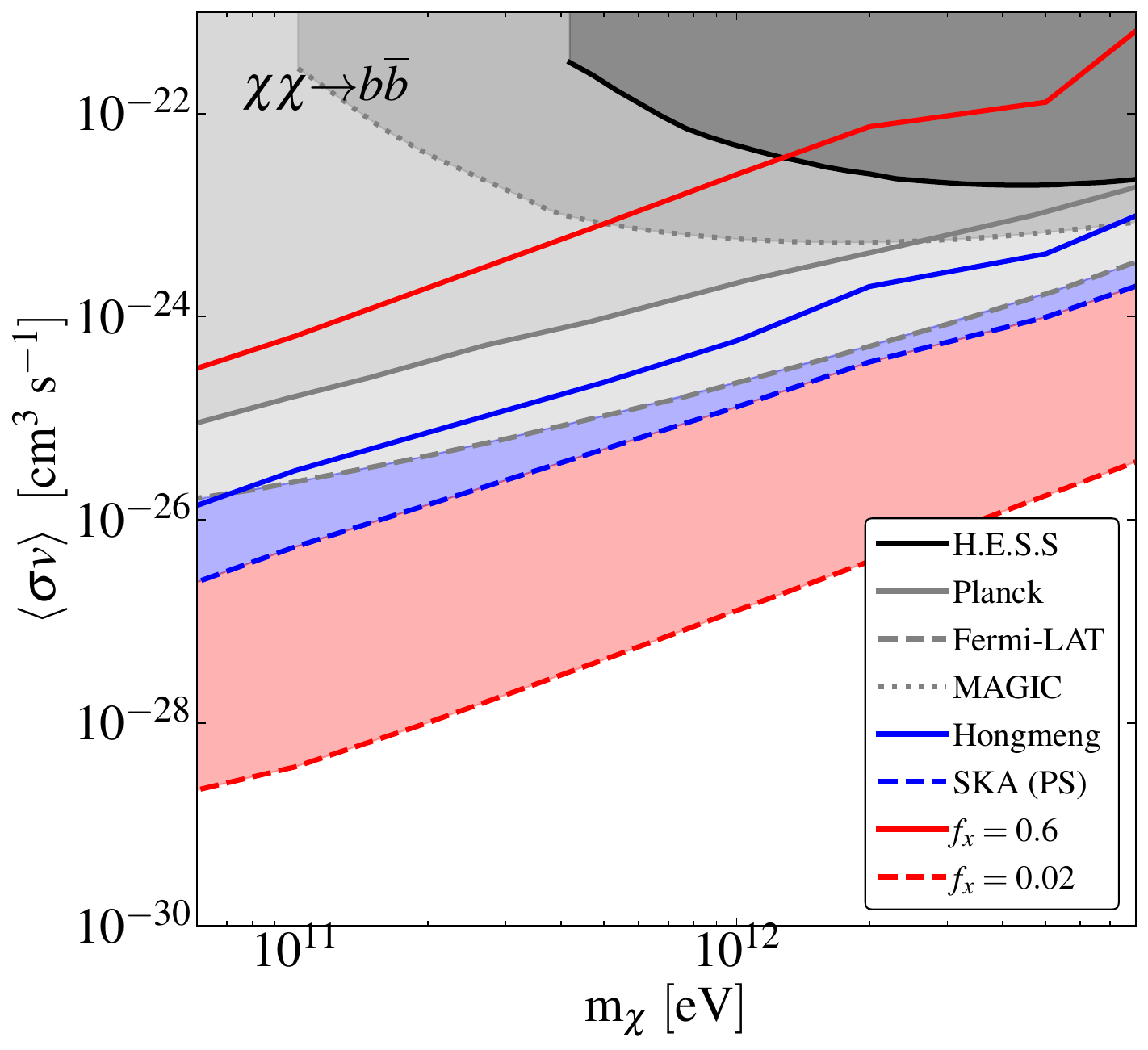}
    \caption{{$2\sigma$ limits on the \ac{DM} annihilation cross section from the \ac{SKA} 21 cm forest \ac{1D} power spectrum.
    The left, middle, and right panels show the constraints for the
    electron-positron pairs, photon pairs, and bottom-anti-bottom quark pairs annihilation channels, respectively.
    Our results are shown by the red curves.
     Also shown for comparison are existing $2\sigma$ upper limits from observations of \ac{CMB} distortion \cite{Zhang:2023usm}, gamma-ray observations \cite{Cirelli:2020bpc,HESS:2018kom,HESS:2014zqa,VERITAS:2017tif,MAGIC:2017avy,Aleksic:2013xea,Fermi-LAT:2015att}, and electron-positron pairs \cite{Cohen:2016uyg,Boudaud:2018oya}, along with constraints from the 21 cm global signals \cite{ zhao:2024jad} and the 21 cm power spectrum \cite{zhao:2025ddy}.
     The shaded regions correspond to the excluded parameter space.}
    }
    \label{fig:anni_forest}
\end{figure*}

\begin{figure*}[!htbp]
    \centering
    \includegraphics[width=0.32\linewidth]{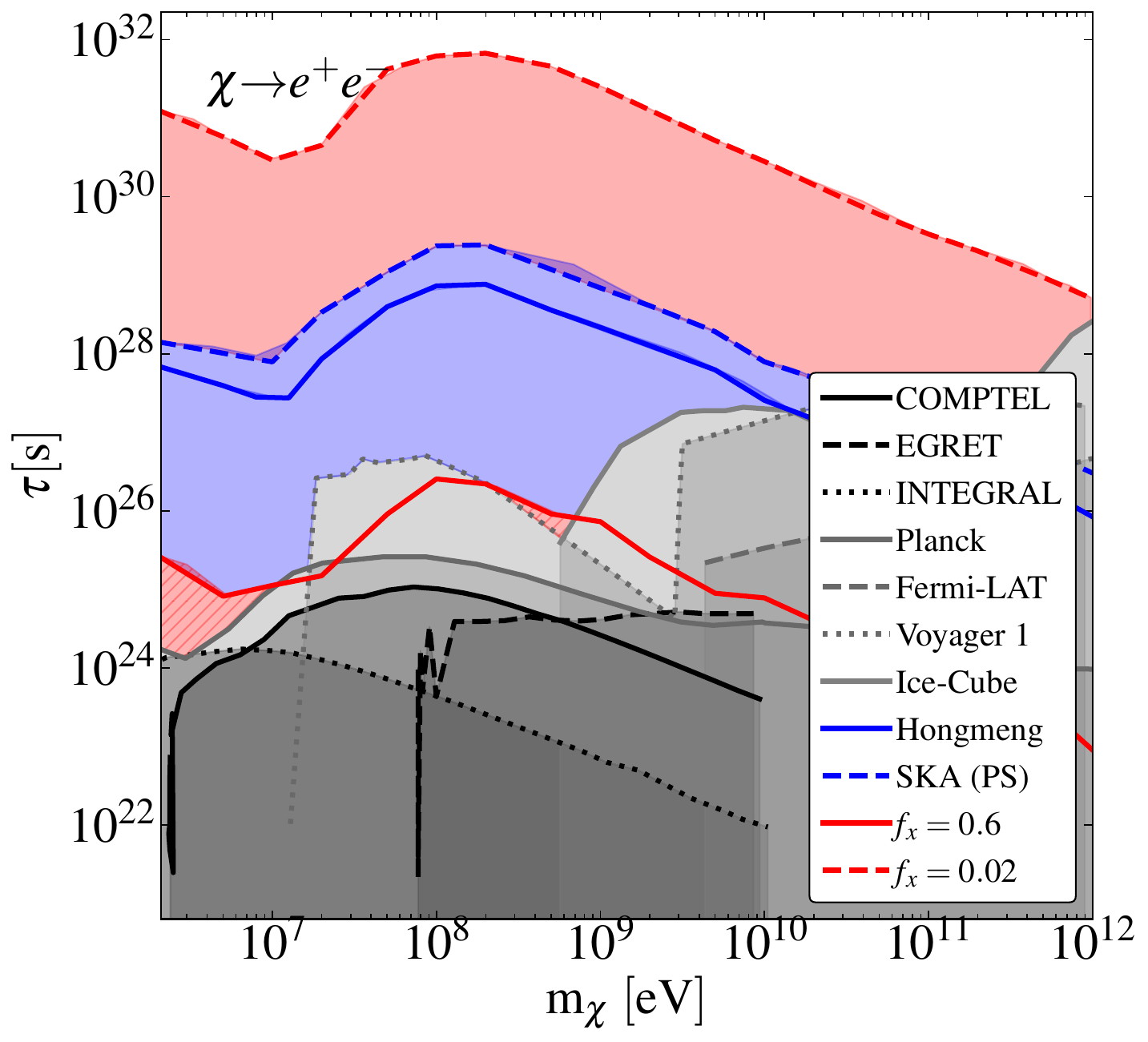}
    \includegraphics[width=0.32\linewidth]{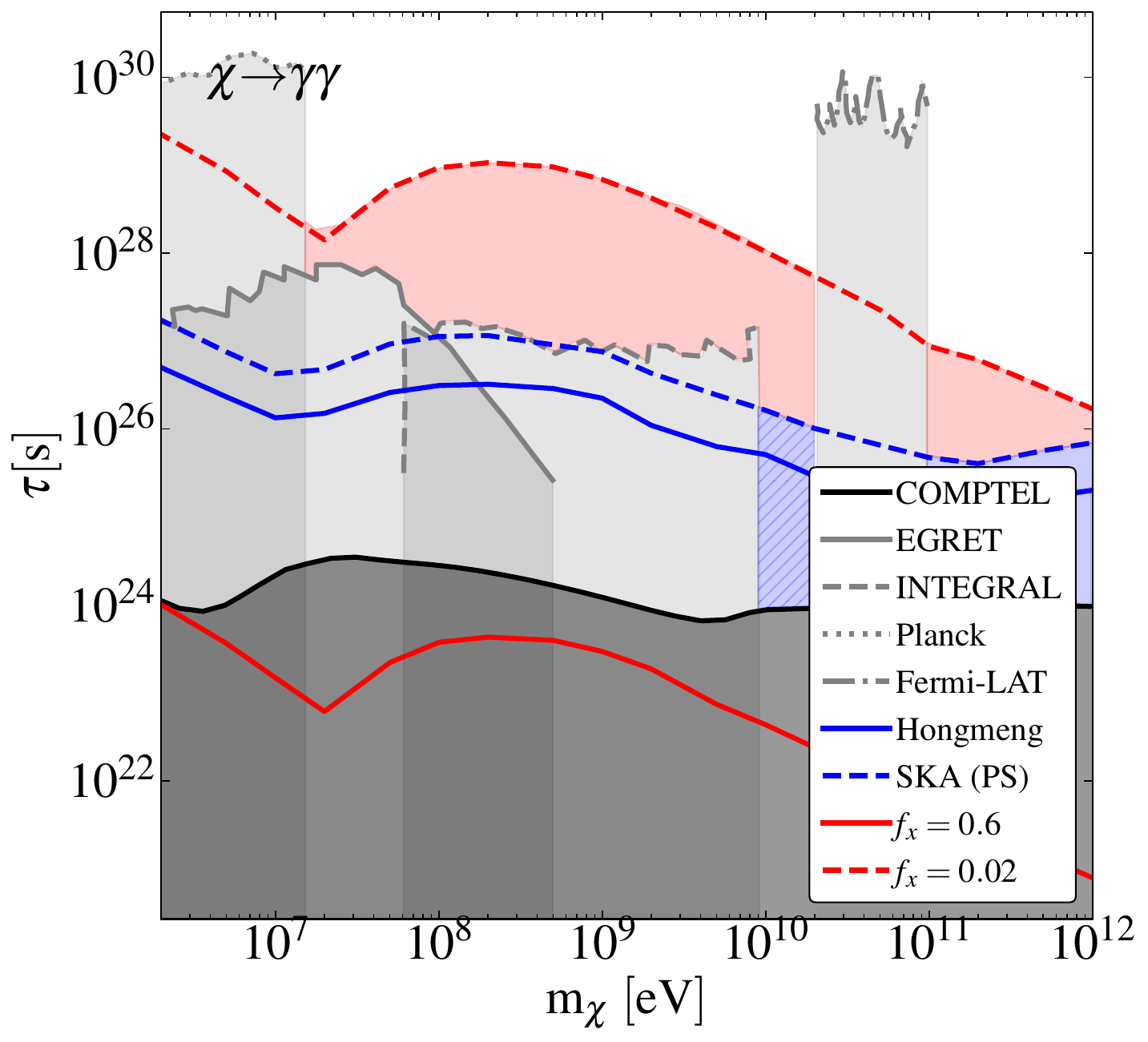}
    \includegraphics[width=0.32\linewidth]{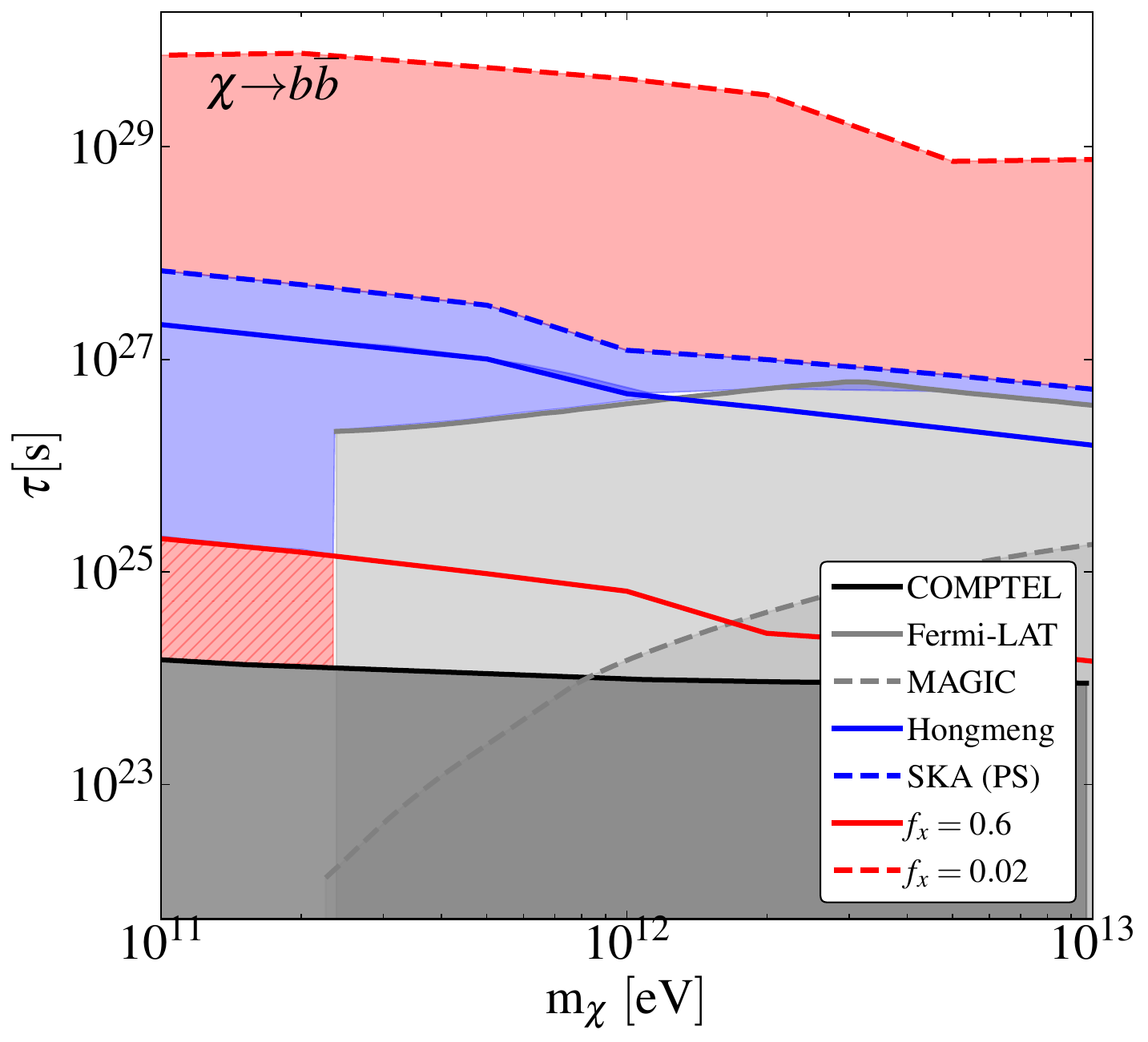}
    \caption{
    {$2\sigma$ limits on the \ac{DM} decay lifetime from the \ac{SKA} 21 cm forest \ac{1D} power spectrum.
    The left, middle, and right panels show the constraints for the
    electron-positron pairs, photon pairs, and bottom-anti-bottom quark pairs decay channels, respectively.
    Our results are shown by the red curves.
    Also shown for comparison are existing $2\sigma$ upper limits from observations of \ac{CMB} distortion \cite{Planck:2018vyg,Capozzi:2023xie}, extragalactic photons \cite{Koechler:2023ual,Calore:2022pks,Foster:2022nva,Cirelli:2020bpc,Massari:2015xea,Cohen:2016uyg,Essig:2013goa}, and electron-positron pairs \cite{Cohen:2016uyg,Boudaud:2018oya}, along with constraints from the 21 cm global signals \cite{ zhao:2024jad}, and the 21 cm power spectrum \cite{zhao:2025ddy}.
    The shaded regions correspond to the excluded parameter space. To maintain visual clarity, areas that are fully overlapped by other shading are not explicitly indicated.}
    }
    \label{fig:decay_forest}
\end{figure*}

\begin{figure}
    \centering
    \includegraphics[width=1.0\linewidth]{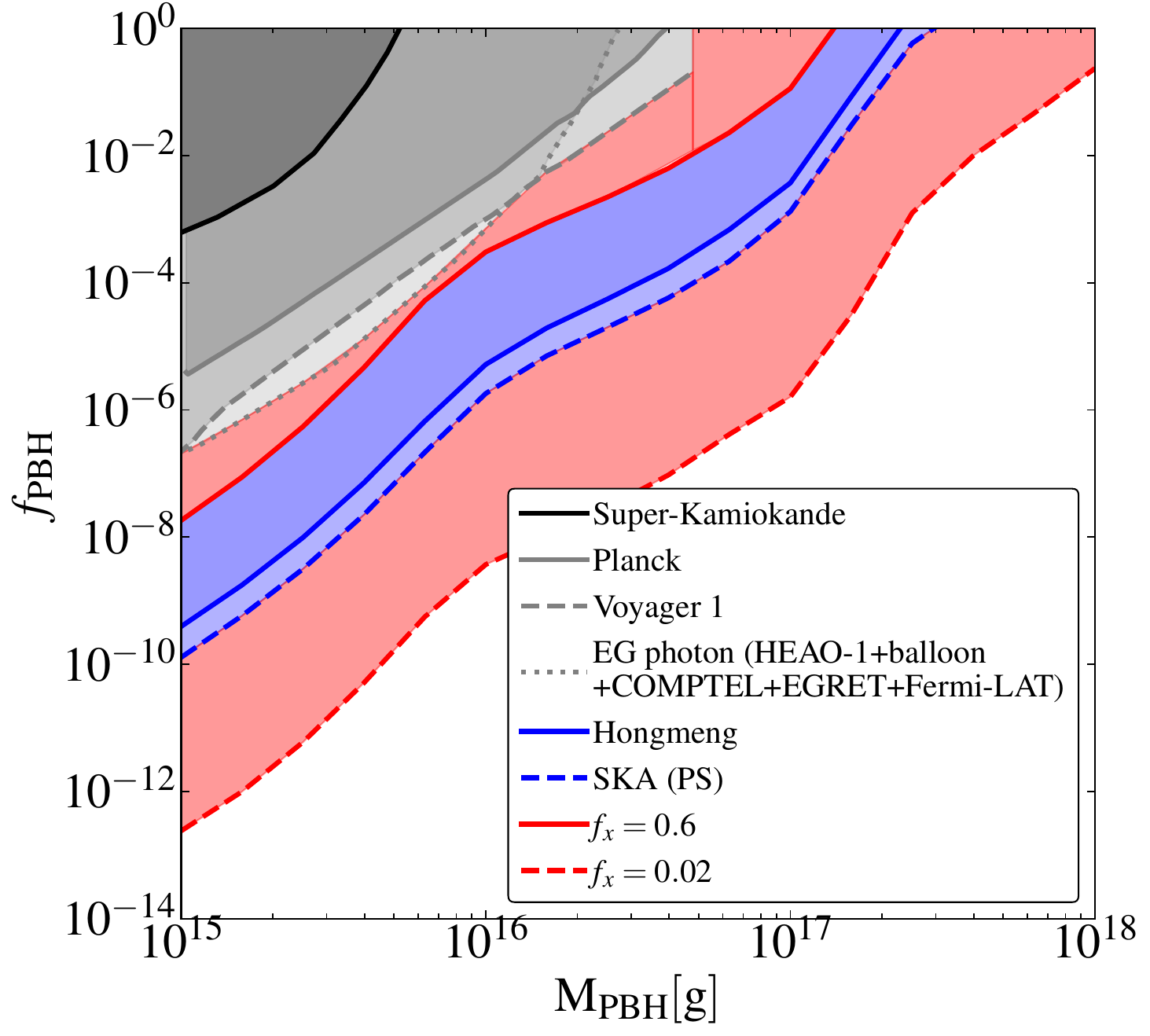}
    \caption{
    {$2\sigma$ limits on the \ac{PBH} mass function from the \ac{SKA} 21 cm forest \ac{1D} power spectrum.
        Our results are shown by the red curves.
     For comparison, we show the existing upper limits at $2\sigma$ confidence level from observations of the diffuse neutrino background \cite{Wang:2020uvi}, \ac{CMB} anisotropies \cite{Chluba:2020oip,Acharya:2020jbv,Clark:2016nst}, extragalactic photons \cite{Carr:2016hva}, and electron-positron pairs \cite{Boudaud:2018hqb}, along with constraints from the 
     21 cm global signals \cite{ zhao:2024jad}, and the 21 cm power spectrum \cite{zhao:2025ddy}.The shaded regions correspond to the excluded parameter space. }}
    
    \label{fig:bh_forest}
\end{figure}

The results of our Fisher matrix analysis are summarized in Figs.~\ref{fig:ann_correlation_05}--\ref{fig:bh_forest}.
Figs.~\ref{fig:ann_correlation_05} and \ref{fig:ann_correlation_002} illustrate the correlations among the model parameters and the constraints on them, assuming a \ac{DM} particle mass of $100~\rm\,MeV$, a \ac{PBH} mass of $10^{16}\,\rm g$, and an observation time of $1000\,\rm hours$.
Figs.~\ref{fig:anni_forest}--\ref{fig:bh_forest} present the prospective $2\sigma$-confidence-level sensitivity of the \ac{SKA} for probing the annihilation and decay of \ac{DM} particles, as well as Hawking radiation from \acp{PBH}, using the 21 cm forest \ac{1D} power spectrum.
For comparison, we include existing upper limits at the $2\sigma$ confidence level from observations of the \ac{CMB} (Planck) \cite{Clark:2016nst,Planck:2018vyg,Chluba:2020oip,Acharya:2020jbv,Capozzi:2023xie,Zhang:2023usm}, gamma rays (\ac{H.E.S.S}, \ac{VERITAS}, \ac{MAGIC}, \ac{Fermi-LAT}, etc.) \cite{Essig:2013goa,Aleksic:2013xea,HESS:2014zqa,Massari:2015xea,Fermi-LAT:2015att,Dong:2015yjs,Cohen:2016uyg,Carr:2016hva,MAGIC:2017avy,VERITAS:2017tif,HESS:2018kom,Cirelli:2020bpc,Calore:2022pks,Foster:2022nva,Koechler:2023ual,Dai:2024guo,Khan:2025kag,Wang:2025jhy}, and electron-positron pairs (Voyager-1) \cite{Boudaud:2016mos,Boudaud:2018oya,Boudaud:2018hqb}.
We also compare our results to the constraints obtained from the 21 cm global spectrum and the 21 cm power spectrum. 


Figs.~\ref{fig:ann_correlation_05} and \ref{fig:ann_correlation_002} reveal a strong degeneracy between X-ray productivity and \ac{DM} parameters on the 21 cm forest \ac{1D} power spectrum.
This strong degeneracy originates from the intrinsic nature of the 21 cm forest signal.
Consequently, the 21 cm forest \ac{1D} power spectrum alone cannot simultaneously and independently constrain both the astrophysical X-ray heating and the \ac{DM}-induced heating effects.
Therefore, to constrain \ac{DM} parameters using the 21 cm forest signal, it is necessary to incorporate other probes to precisely determine the astrophysical X-ray heating.

Fig. \ref{fig:anni_forest}
shows that, 21 cm forest \ac{1D} power spectrum performs better in probing \ac{DM} particle annihilation via $\chi\chi\rightarrow e^{+}e^{-}$ channel compared to other channels such as $\chi\chi\rightarrow\gamma\gamma$ and $\chi\chi\rightarrow b \bar{b}$.
Focusing on the optimal annihilation channel $\chi\chi \rightarrow e^{+}e^{-}$ (left panel), we find that utilizing the \ac{SKA}, with $1,000$ hours of integration time and $f_{\rm X} = 0.02$ (shown by the red dashed curve), the 21 cm forest \ac{1D} spectrum can achieve a sensitivity of $\langle\sigma v\rangle \leq 10^{-31}\,{\rm cm}^{3}\,{\rm s}^{-1}$ for $10$\,GeV \ac{DM} particles.
This constraint surpasses the most stringent existing constraints (gray curves) by up to $4$ orders of magnitude.
This result indicates that, under low astrophysical heating, the 21 cm forest signal  could improve upon existing limits from other observations in the near future.
However, as indicated by the red solid curve, if astrophysical heating is very strong, the 21 cm forest signal cannot provide competitive constraints.
These results demonstrate that the ability to constrain \ac{DM} parameters using the 21 cm forest signal is critically dependent on the level of astrophysical X-ray heating.
Therefore, it is necessary to incorporate independent constraints on the astrophysical X-ray heating efficiency to effectively constrain \ac{DM} parameters using the 21-cm forest signal.

We also compare our constraints on \ac{DM} annihilation from the 21-cm forest 1D power spectrum with those from our previous work \cite{ zhao:2024jad,zhao:2025ddy}.
The results of our previous work, which constrain \ac{DM} properties utilizing the 21 cm global spectrum and the 21 cm power spectrum, are shown by the blue curves.
By comparing the red and blue curves, we find that, under low astrophysical heating scenarios, the 21 cm forest \ac{1D} power spectrum can achieve constraints up to $2$ orders of magnitude tighter than those from the 21 cm global spectrum and the 21 cm power spectrum with the same observation time.
However, this requires a precise measurement of astrophysical heating processes, which reemphasizes the importance of incorporating independent constraints on the astrophysical X-ray heating efficiency.

Based on the results of Fig. \ref{fig:decay_forest}, we find that the 21 cm forest \ac{1D} power spectrum has superior sensitivity for probing \ac{DM} decay through the $\chi \rightarrow e^{+}e^{-}$ channel, compared to alternative channels such as $\chi \rightarrow \gamma\gamma$ and $\chi \rightarrow b\bar{b}$.
Focusing on this optimal channel $\chi \rightarrow e^{+}e^{-}$ (left panel) and utilizing the \ac{SKA} with $1,000$ hours of integration time under low astrophysical heating scenarios (red dashed curve), we find that the 21 cm forest \ac{1D} power spectrum achieves a sensitivity of $\tau \geq 10^{30}$\, seconds for $10$\,GeV \ac{DM} particles.
This result is more stringent than the most stringent existing constraints (gray curves) by up to $4$ orders of magnitude.
This indicates that the 21 cm forest signal could test existing limits on \ac{DM} particle decay from other observations in the near future.
Furthermore, in the sub-GeV range, the 21 cm forest \ac{1D} power spectrum has the potential to improve the existing constraints from other probes by up to six orders of magnitude under low astrophysical heating scenarios.
However, the sensitivity is affected by astrophysical heating, as shown by the red solid curve.
If astrophysical heating is strong, the 21 cm forest \ac{1D} power spectrum cannot provide competitive constraints.
This underscores the importance of precise independent measurements of astrophysical processes using other probes.
Therefore, collaborative observations with other facilities represent an important direction for future 21 cm forest research.
Moreover, we compare the constraints from the 21 cm forest \ac{1D} power spectrum with those from our previous work, which constrained \ac{DM} particle decay using the 21 cm global spectrum and 21 cm power spectrum (shown by the blue curves). 
We find that only under low astrophysical heating scenarios is the 21 cm forest \ac{1D} power spectrum able to achieve constraints better than those obtained from the 21 cm global spectrum and the 21 cm power spectrum with the same integration time, similar to the case for \ac{DM} particle annihilation.

Figure \ref{fig:bh_forest} demonstrates that the 21 cm forest \ac{1D} power spectrum measured by the \ac{SKA} achieves a sensitivity to $f_{\rm PBH} \simeq 10^{-13}$ for \acp{PBH} with masses of $10^{15}$\,g with an integration time of $1,000$\,hours under low astrophysical heating scenarios.
This result surpasses constraints from existing observations (gray curves) by up to $5-6$ orders of magnitude, indicating that the \ac{SKA} can improve upon these limits in the near future. 
Moreover, the \ac{SKA} can probe higher-mass \acp{PBH} compared to current experiments, particularly probing uncharted parameter space above $10^{18}$\,g.
However, under strong astrophysical heating scenarios, the 21 cm forest \ac{1D} power spectrum provides only constraints comparable to existing limits.
This necessitates independent measurements of astrophysical processes via other probes.

\section{Conclusions}
\label{sec:summary}

This study has investigated the potential of the 21 cm forest 1D power spectrum as a novel probe of \ac{DM} physics, emphasizing the \ac{SKA}'s capacity to detect signatures of \ac{DM} annihilation, decay, and \ac{PBH} evaporation during cosmic reionization. We have shown that these processes inject excess energy into the \ac{IGM}, modifying the thermal history of the early universe and imprinting distinctive features on the 21 cm forest 1D power spectrum. Through detailed modeling and sensitivity analysis, we have quantified the \ac{SKA}'s potential to constrain \ac{DM} parameters.
Our results have shown that, in the near future, the 21 cm forest 1D power spectrum has the potential to serve as a powerful tool for small-scale \ac{DM} investigations.
Additionally, if combined with other astrophysical probes, the 21 cm forest 1D power spectrum is expected to provide valuable insights into both astrophysical processes and \ac{DM} properties.

The 21 cm forest signal, observed with the \ac{SKA} under low astrophysical heating scenarios, can constrain \ac{DM} annihilation and decay.
For annihilation channels producing $e^+e^-$ pairs, in this case, the 21 cm forest \ac{1D} power spectrum achieves constraints up to $4$ orders of magnitude tighter than current experimental limits.
Similarly, for decay channels yielding $e^+e^-$ pairs, it can improve upon existing sensitivity limits by up to $4$ orders of magnitude.
Moreover, the 21 cm forest signal is exceptionally sensitive to sub-GeV \ac{DM} particles, exceeding current detection limits.
These findings establish the 21 cm forest \ac{1D} power spectrum as a powerful probe of \ac{DM} on small scales.
However, constraints on \ac{DM} from the 21 cm forest \ac{1D} power spectrum are strongly dependent on astrophysical heating processes.
Under strong astrophysical heating scenarios, this method fails to provide competitive constraints.
Therefore, it is essential to break the degeneracy by precise determination of astrophysical heating via independent probes.  
Additionally, sensitivity can be enhanced by extending integration time or by observing a larger sample of radio-bright background sources, thereby strengthening \ac{DM} parameter constraints.

The \ac{SKA} has unique potential for probing \acp{PBH} through the 21 cm forest \ac{1D} power spectrum.
Our analysis establishes a constraint on $f_{\rm PBH}\simeq10^{-13}$ for \acp{PBH} with masses of  $10^{15}$\,g under low astrophysical heating scenarios, surpassing current limits by $5-6$ orders of magnitude.
Critically, utilizing the 21 cm forest \ac{1D} power spectrum, the \ac{SKA} extends these constraints to the high-mass regime of $10^{18}$\,g , a parameter space unexplored by conventional detection methods.
To further improve sensitivity and broaden the applicable mass range, expanding the observed quasar number and extending integration times are essential.

In summary, our analysis demonstrates that under low astrophysical heating, the 21 cm forest \ac{1D} power spectrum can effectively probe \ac{DM} annihilation, decay, and \ac{PBH} Hawking radiation during the epoch of reionization, using \ac{SKA} with $1000$ hours of integration time.
However, under strong astrophysical heating, the 21 cm forest signal cannot provide competitive constraints.
In this work, we focus on the impact of telescope thermal noise and sample variance on the 21 cm forest signal, which manifests as a series of absorption lines in the spectra of radio background sources. Analyzing this signal requires the removal of the continuum component. This process simultaneously eliminates smoothly distributed foreground emission, resulting in relatively low foreground contamination. Additional noise sources, such as telescope beam effects and radio frequency interference, also play a significant role in real observations and must be addressed in future research.

While this study focuses exclusively on the potential of the 21 cm forest \ac{1D} power spectrum for probing \ac{DM} using \ac{SKA} under various scenarios, future efforts adopting multi-scale approaches combining complementary probes (21 cm global spectrum, 21 cm power spectrum, and 21 cm forest signal) will advance our understanding of early-universe physics. 
Specifically, the 21 cm global spectrum and 21 cm power spectrum can provide precise measurements of astrophysical heating processes.
When combined with the 21 cm forest signal, this multi-probe synergy will provide unprecedented constraints on \ac{DM} properties in the early universe.
Moreover, by systematically exploring \ac{DM} across spatial scales from full-sky (21 cm global spectrum) to large scales (21 cm power spectrum) and down to small scales (21 cm forest signal), we can holistically characterize its nature.
This integrated methodology will ultimately deliver powerful observational constraints on \ac{DM} processes including annihilation, decay, and Hawking radiation during the early universe.
In summary, 21 cm cosmology holds exceptional promise as a powerful probe for unveiling both the physics of the early universe and the fundamental nature of \ac{DM}.

\section{DATA AVAILABILITY}
The data that support the findings of this article are openly available \cite{Zhao_data}.

\section*{Acknowledgments}
We thank Yidong Xu for the helpful discussion and also for providing the code for the 21 cm forest multi-scale simulations. 
This work is supported by the National Natural Science Foundation of China (Grant Nos. 12533001, 12473001, and 12175243), the National SKA Program of China (Grant Nos. 2022SKA0110200, 2022SKA0110203), the National Key R\&D Program of China (Grant No. 2023YFC2206403), the China Manned Space Program (Grant No. CMS-CSST-2025-A02), and the 111 Project (Grant No. B16009).

\bibliography{main}

\end{document}